\documentclass{article}
\usepackage{spconf,amsmath,graphicx}      
\usepackage{comment}
\usepackage{xcolor}

\title{Computer Vision Methods for Frequency Analysis \\ of RFI in Radio Astronomy Data}
\name{Natalia A. Schmid, Sasanka Katreddi, and Yechan Kweon
\thanks{This research is supported by the National Science Foundation under Award No. AST-2307581. The authors would also like to thank their colleagues at the Green Bank Observatory and West Virginia University for providing the data set used throughout this research.
}}
\address{Lane Department of Computer Science and Electrical Engineering, \\ West Virginia University, Morgantown, WV, USA.}
\begin{document}
%
\maketitle
\begin{abstract}
Radio Frequency Interference (RFI) increasingly contaminates the radio astronomy spectrum, often exceeding astronomical signal amplitudes by 50–70 dB. Reliable detection and mitigation are therefore essential for studies of faint transient phenomena such as pulsars and fast radio bursts (FRBs). Existing practical methods (including Spectral Kurtosis (SK), Median Absolute Deviation (MAD), and SumThreshold) perform well in many settings but depend on assumptions about the RFI environment and data statistics, limiting their effectiveness for weak, broadband, or non stationary interference. 

We develop a transform based RFI detection method that requires no prior knowledge of RFI origin or type. Using Green Bank Telescope (GBT) data containing PSR J1713+0747, with 4096 channels spanning 1.1–1.9~GHz and 5.12~$\mu$s sampling, we apply a Short Time Fourier Transform (STFT) to each channel and use an image segmentation algorithm on the STFT magnitude to generate a binary RFI mask. The masked data are inverse transformed and reassembled into a cleaned time series. 

Performance is assessed using the Signal to Noise Ratio (S/N) of a single pulse of PSR J1713+0747, with SK serving as the baseline. The cleaned spectrogram is dedispersed, integrated across frequency, and evaluated through the resulting S/N.  Experimental results show that refining each channel's frequency content via STFT, followed by segmentation in the STFT domain, yields measurable improvements in RFI suppression. 

\end{abstract}
\section{Introduction}
\label{sec:intro}
The frequency spectrum allocated to radio astronomy is increasingly compromised by powerful and sophisticated terrestrial radio transmissions, limiting the ability of astronomers to probe the universe at greater depth. In many cases, the amplitude of RFI exceeds that of astronomical signals by 50–70 dB, making effective mitigation essential for producing clean and scientifically reliable data.

Accurate detection and classification of RFI are therefore critical preprocessing steps before searching for faint astrophysical phenomena such as pulsars and FRBs. Although manual labeling can be highly accurate, it is prohibitively time‑consuming and infeasible for modern large‑scale datasets. Consequently, recent years have seen rapid growth in classical machine‑learning techniques \cite{Rousseeuw1993MAD,McLachlan2000Mixtures,nita2007,nita2010,Offringa2010PostCorrelation}, statistical and information‑theoretic approaches \cite{Pinto2015EntropyRFI,Offringa2013LOFAR,Tikhonov2015DivergenceRFI,Morales2015KLD}, and unsupervised or weakly supervised deep‑learning architectures \cite{An2015VAE,Schlegl2017GANAnomaly,Baur2021Autoencoders} tailored to RFI identification and excision. These efforts have produced a wide range of algorithms implemented in both software and hardware, substantially improving the efficiency and robustness of radio‑astronomical data processing. 

Among the most widely used methods for RFI mitigation are the Spectral Kurtosis (SK) estimator \cite{nita2007,nita2010,nita2016}, Median Absolute Deviation (MAD) thresholding \cite{Rousseeuw1993MAD,Real-timeRFI:eramey}, and the SumThreshold algorithm \cite{Offringa2010PostCorrelation}. While these methods are effective in many observing contexts, their performance depends on assumptions about the local RFI environment and the statistical properties of the data. As a result, they may respond differently to weak, broadband, or non‑stationary interference and may fail to capture the full range of RFI morphologies encountered across instruments. These considerations motivate the use of complementary approaches, such as the transform‑based method introduced here, that can adapt more flexibly to diverse data characteristics.

\begin{figure*}[ht]
    \centering
    \includegraphics[width=0.85\textwidth]{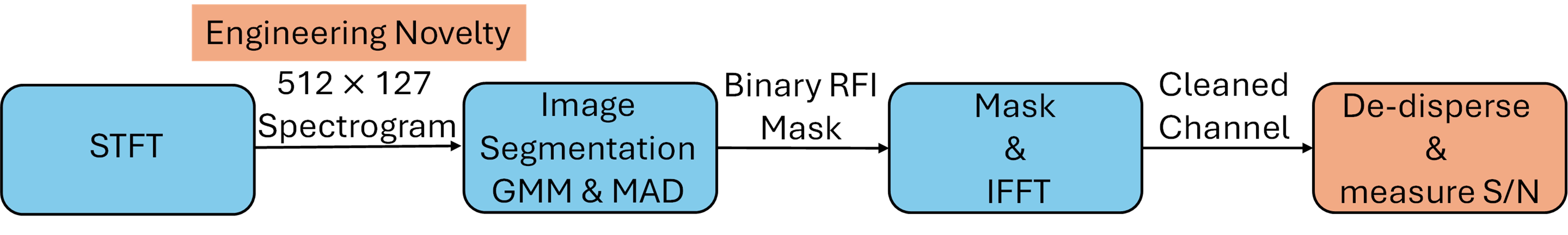}
    \caption{A block diagram summarizing the main steps of the proposed RFI‑mitigation pipeline.}
    \label{fig:block_diagram}
\end{figure*}
In this work, we investigate the removal of RFI from data containing astronomical transient signals such as FRBs and pulsars. Upon arrival at a radio telescope, these transients are typically attenuated, dispersed by the ionized Interstellar Medium (ISM), and embedded within substantial thermal noise and RFI. 

A defining characteristic of dispersed transient signals is the frequency‑dependent arrival time imposed by propagation through the cold, ionized ISM. Higher-frequency components reach the telescope earlier than lower-frequency components. When the raw voltage data are converted into a time–frequency representation via a short‑time Fourier transform (STFT), a transient, if detectable above the noise and RFI, appears as a downward‑sweeping feature across the spectrogram. 

The relative delay between two observing frequencies $f_1$ and $f_2$ is given by the standard cold‑plasma dispersion relation: 
\begin{equation}
\Delta t\; =\; 4.148808\times 10^3\; \mathrm{ms}\; \left( \frac{1}{f_1^2}\; -\; \frac{1}{f_2^2}\right) \left( \frac{\mathrm{DM}}{\mathrm{pc\cdot cm^{-3}}}\right),
\label{eq:dispersion_equation}
\end{equation} 
where $\mathrm{DM}$ is the dispersion measure, defined as the line‑of‑sight integral of the free‑electron density. This quadratic frequency dependence produces the characteristic sweep observed in dynamic spectra of transient radio sources.

The goal of this research is to develop a new transform-based RFI detection method under the assumption that the origin and type of RFI are unknown.  We are working with data from the GBT that contain the known pulsar PSR J1713+0747, along with several known and unknown RFI sources. 

As a performance metric, we will use the Signal-to-Noise Ratio (S/N) value of a single pulse of PSR J1713+0747 contained in the dataset. The SK method, commonly used by radio astronomers for RFI mitigation, will serve as the baseline for performance comparison. 

Unlike other RFI‑mitigation methods that rely solely on frequency‑domain or time‑domain analysis, our approach begins by applying a Short‑Time Fourier Transform (STFT) to {\em each frequency channel} in the original spectrogram to obtain its time–frequency representation. We then apply an image‑segmentation method to the magnitude of the transformed data, producing a binary mask that distinguishes RFI from clean samples. This mask is subsequently applied to the STFT data generated in the initial step. Finally, the RFI‑cleaned data are inverse‑Fourier‑transformed and reassembled into a continuous time series, which replaces the original data in the channel. 

The cleaned data are then processed through a standard radio astronomy pipeline: the clean spectrogram is de-dispersed, then integrated across frequency, and the S/N of the resulting time series is assessed to evaluate the effectiveness of our method. 

Experimental results show that refining each channel’s frequency content via STFT, followed by segmentation in the STFT domain, which are the two key engineering novelties implemented in this work, yield measurable improvements in RFI mitigation.

The remainder of the paper is organized as follows. Section~\ref{sec:methodology} describes the proposed RFI‑mitigation pipeline used in this work. Section~\ref{sec:experimental_analysis} presents the observational and qualitative results for the methods explored in this study, illustrating the RFI‑segmentation capability of the proposed approach and analyzing the resulting S/N values. Finally, Section~\ref{sec:summary} summarizes the main findings.

\section{RFI-mitigation Pipeline}
\label{sec:methodology}
Consider a spectrogram with $K$ discrete frequency channels and $M$ time samples,  formed from real and imaginary parts of complex valued channelized voltages. Let $\Delta W$ be the bandwidth of each frequency channel. For the GBT L-band data $\Delta W$ is equal to $800,000,000/4096 = 195,312.5$ Hz. Because we assume no prior knowledge of the origins of RFI or their operating frequencies, we likewise have no information about which of the K channels, or which sub‑band within a channel’s bandwidth $\Delta W$ is contaminated by RFI. Under the additional assumption that {\em RFI typically appears as high‑energy, structured features within} $\Delta W,$ we adopt the signal‑processing approach outlined in Figure \ref{fig:block_diagram}.

\subsection{Applying STFT}
The original spectrogram with $K$ frequency channels and $M$ time samples is generated from high resolution complex valued voltage data. Here we perform a similar transformation on data that span the $195,312.5$ Hz frequency band. We apply the STFT of window size $L$ to the data in each channel.  The result is a complex valued frequency-time array with $L$ rows and $M/L$ columns assuming that $M$ is multiple of $L.$ Denote this complex valued array as $Z_k,$ with $k$ indicating that STFT was applied to data in channel $k,$ $k = 1,...,K.$     Applying the magnitude to the array yields an image, denote it by $X_k$     

\subsection{Applying an Image Segmentation Method}
In this research, we involve two classical image segmentation approaches to segment RFI in magnitude STFT data per channel: the Gaussian Mixture Model (GMM) \cite{pal1993image} and an Energy-based approach with MAD thresholding \cite{rousseeuw1993alternatives}.

\vskip \medskipamount
\noindent
{\bf GMM segmentation }
is a model‑based data‑clustering approach used for region segmentation in images and spectrograms. The GMM framework treats the value of each pixel in a spectrogram as being generated by a mixture of Gaussian probability density functions (pdf), each with its own parameters corresponding to a particular region class. In our case, the model contains only two Gaussian components: one describing RFI and the other describing clean, RFI‑free data. Thus, each pixel $x_k(i,j)$ in the spectrogram of channel $k$ is modeled as having a distribution 
\[
\pi_0\, \mathcal{N}(\mu_0,\sigma _0^2)+(1-\pi_0)\, \mathcal{N}(\mu_1,\sigma_1^2), 
\]
where $\mu_0,$ $\sigma_0^2,$ and $\pi_0$ denote the mean, variance, and prior probability of the RFI component, $\mu_1,$ $\sigma_1^2,$ and $\pi_1=1-\pi_0$ denote the corresponding parameters for the clean‑data component, and $\cal{N}(\cdot,\cdot)$ is used to denote a Gaussian pdf.  The classification of each pixel $x_k(i,j)$ is based on the Maximum a Posteriori Probability (MAP) rule: a pixel is labeled as RFI if its a posteriori probability under the RFI model exceeds that under the clean‑data model, i.e., if $\pi_0 > \pi_1$ after parameter estimation. The iterative solution to this estimation problem uses the Expectation–Maximization (EM) algorithm \cite{dempster1977em}, which at each iteration updates the five unknown parameters $\pi_0,$ $\mu_0,$ $\mu_1,$ $\sigma_0^2,$ and $\sigma_1^2$ based on the observed data.

\vskip \medskipamount
\noindent
{\bf Energy-based MAD thresholding}
directly applies MAD threshold to each pixel in a spectrogram.  Mathematically it is described as follows. Let the median of a spectrogram $X_k$ be denoted as $Mdn(X_k) = median(X_k)$ and the median of the absolute deviation of $X_k$ from the median $Mdn(X_k)$ be described as 
$MAD(X_k) = median(|X_k-Mdn(X_k)|).$  Then the energy-based MAD thresholding rule labels the pixel $x_k(i,j)$ as RFI if 
\begin{equation} 
|x_k(i,j) - Mdn(X_k)| > 1.4828 \cdot MAD(X_k) \cdot \gamma, 
\label{eq:MAD_rule}
\end{equation}
where $\gamma$ is a design parameter typically set to be between 3 and 7.  
 
\subsection{Applying RFI Mask to raw STFT data}
Both the GMM segmentation and the energy‑based MAD thresholding label RFI‑contaminated pixels in the spectrogram of each frequency channel. After segmentation, the result is a binary mask of zeros and ones. A zero in the $(i,j)$-th position indicates that the pixel $x_k(i,j)$ is RFI‑contaminated and should be flagged, whereas a one indicates that the pixel is retained.    

This binary mask is then multiplied point‑by‑point with the matrix $Z_k,$ suppressing contaminated regions while preserving the astronomical signal content.

\subsection{Applying a standard pulsar search pipeline}
After identifying and suppressing the contaminated frequency bins in the STFT domain, the cleaned data are inverse‑transformed back into one‑dimensional time series and reinserted into the original spectrogram, replacing the RFI contaminated samples in each channel. The cleaned spectrogram is then processed through a standard radio‑astronomy pipeline: the data are de-dispersed according to (\ref{eq:dispersion_equation}) and integrated across frequency. Finally, we compute the S/N of the resulting time series to assess the effectiveness of our method. The S/N of a single folded pulse is traditionally used as a measure of signal quality when searching for pulsars \cite{Lorimer12}.

The SK estimator, widely employed in radio astronomy for real‑time RFI excision, is adopted as the baseline against which our method is evaluated.  

SK is a measure of the degree of non‑stationarity and non‑Gaussianity in a signal.  Although the estimator was originally developed for the analysis of SONAR data, Nita et al. \cite{nita2007} introduced the method to the radio astronomy community, where it has since been extended in several forms \cite{nita2010, nita2016, gary2010, nita2019, Taylor2019}.

In practice, SK is applied to data represented as a spectrogram. For a given frequency channel containing $M$ channelized power measurements $P(m),$ departures from unity beyond analytically derived thresholds are identified as outliers. The statistic is computed from the quantities 
\[ 
S_1=\sum _{m=1}^MP(m),\qquad S_2=\sum _{m=1}^MP^2(m), \]
which are combined to form 
\begin{equation}
\mathrm{SK}=\frac{M+1}{M-1}\left( \frac{MS_2}{S_1^2}-1\right) . 
\label{eq:generalizedSK} 
\end{equation} 
The resulting value is compared against a threshold, commonly taken to be $\pm 3\sigma \approx \pm 6/\sqrt{M}.$ Channels for which $\mathrm{SK}$ exceeds this threshold are flagged as contaminated by RFI.

\section{Experimental Results}
\label{sec:experimental_analysis}
%
Prior to the performance analysis, we summarize the observational data used in our experiments. 
\begin{figure*}[ht]
\begin{minipage}[b]{0.49\linewidth}
  \centering
  \centerline{\includegraphics[width=8.5cm]{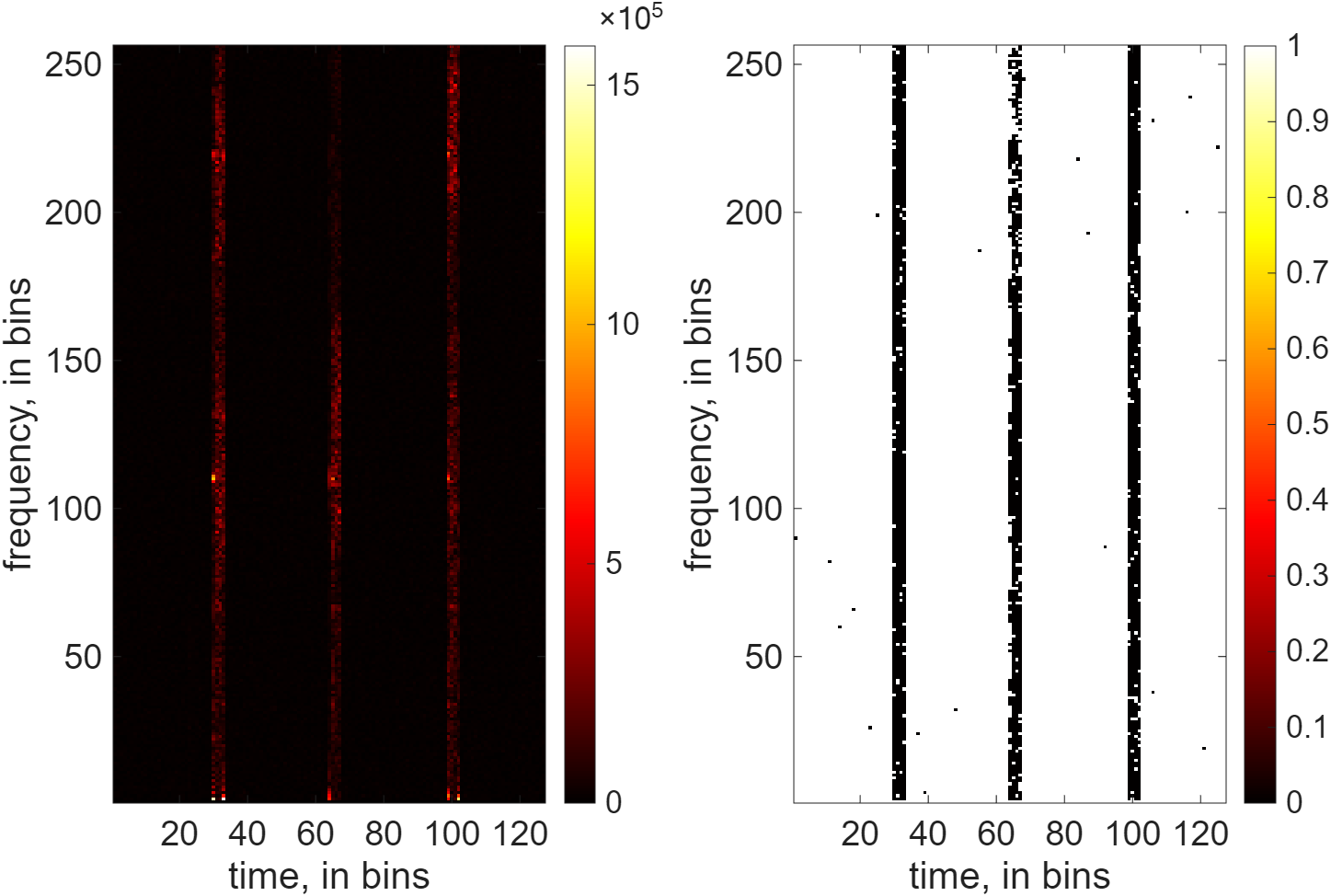}}
  \centerline{(a) Channel 1436}\medskip
\end{minipage}
\begin{minipage}[b]{0.49\linewidth}
  \centering
  \centerline{\includegraphics[width=8.5cm]{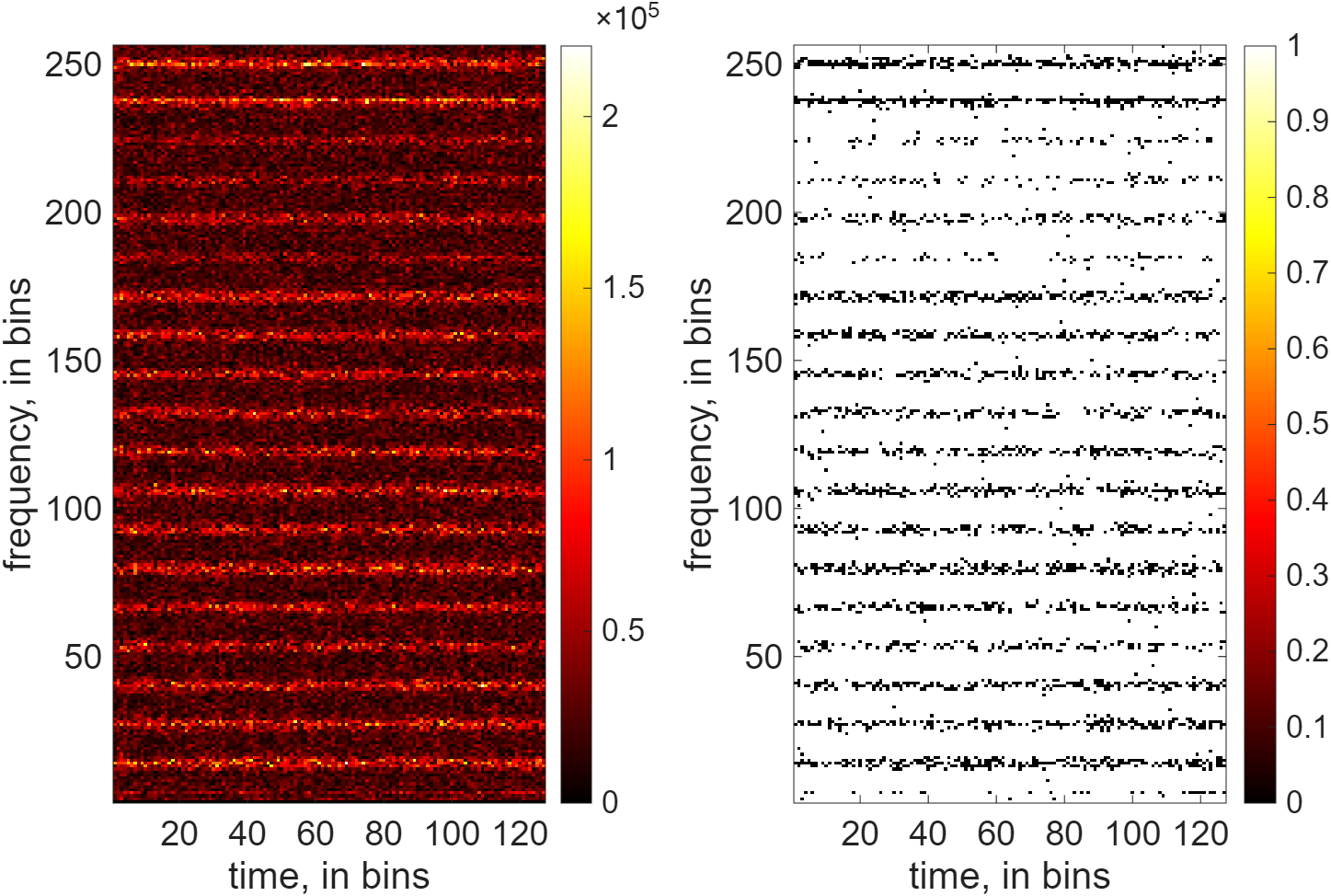}}
  \centerline{(b) Channel 1818}\medskip
\end{minipage}    
\begin{minipage}[b]{0.49\linewidth}
  \centering
  \centerline{\includegraphics[width=8.5cm]{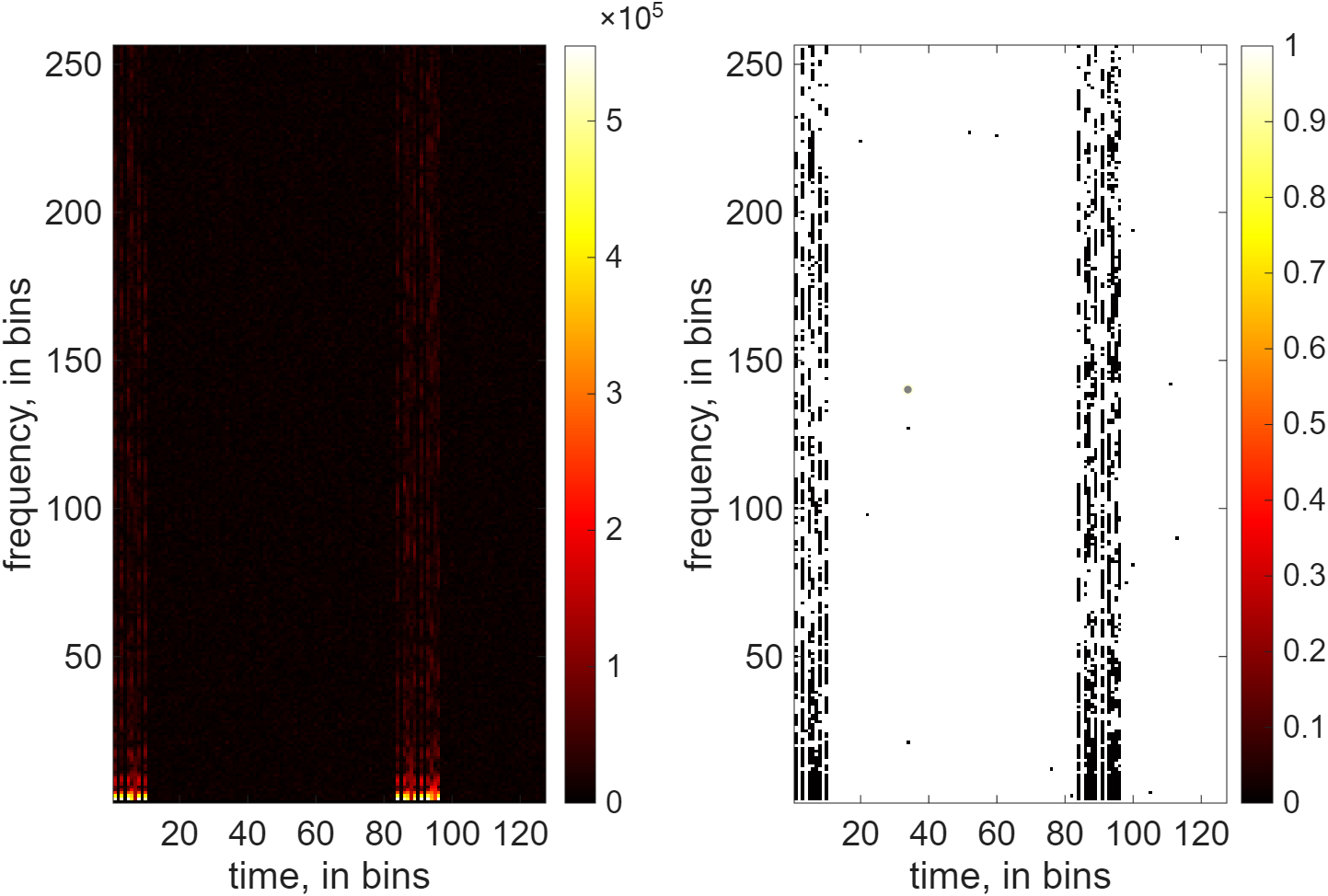}}
  \centerline{(c) Channel 1830}\medskip
\end{minipage}    
\hfill
\begin{minipage}[b]{0.49\linewidth}
  \centering
  \centerline{\includegraphics[width=8.5cm]{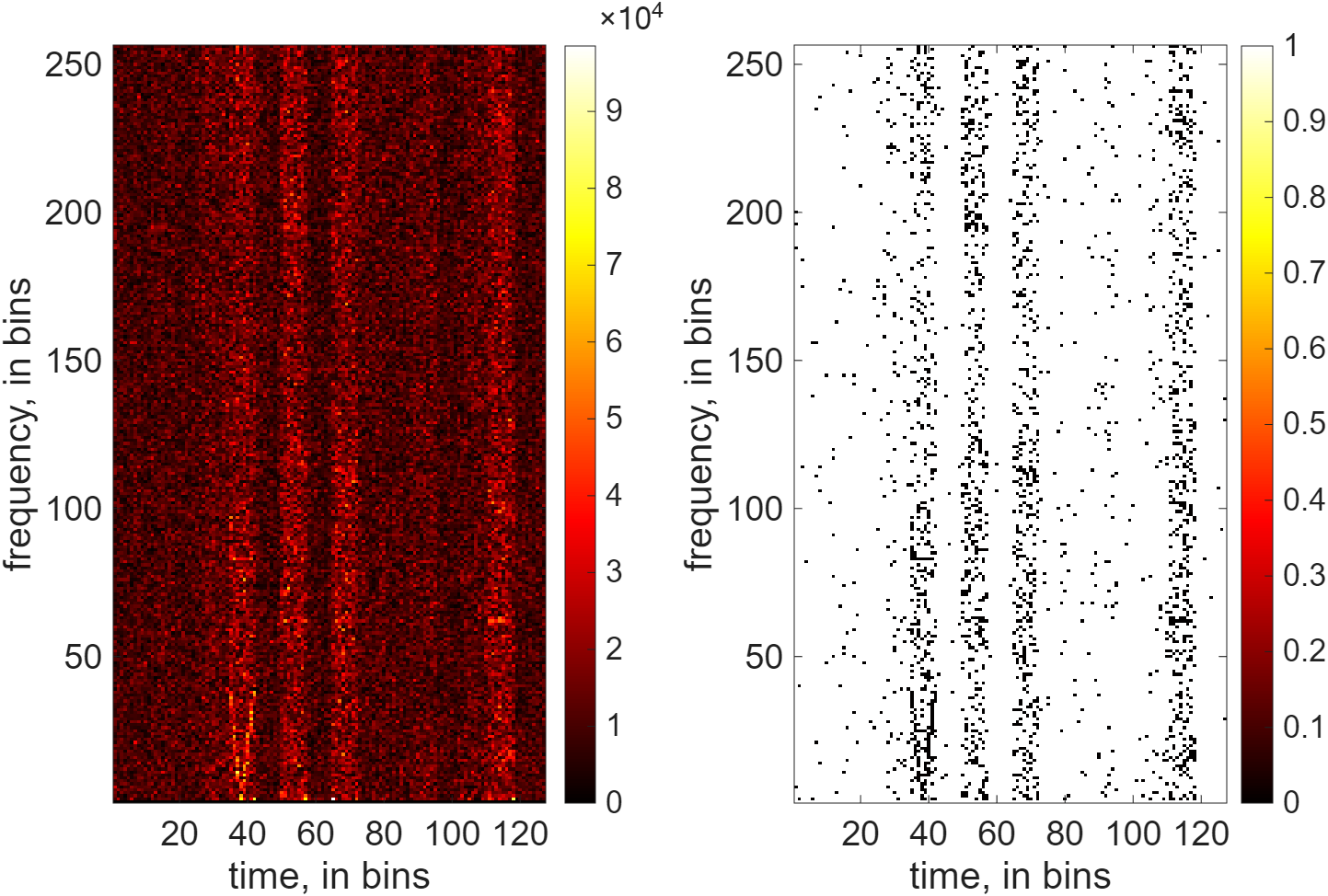}}
  \centerline{(d) Channel 1913  }\medskip
\end{minipage}
\caption{ These examples show segmentation performance across a range of RFI morphologies. Regardless of whether the interference is narrowband, time-varying, or densely structured, the segmentation algorithms successfully separate RFI from the background noise floor.}
\label{fig:res}
\end{figure*}

\subsection{Data} 
The dataset spans an 800~MHz band channelized into 4096 non‑overlapping bins (Channel~4096~$\longrightarrow$~1100~MHz; Channel~1~$\longrightarrow$~1900~MHz). Raw data consist of complex voltages in two orthogonal polarizations, and PSDs are formed from their squared magnitudes.
RFI‑mitigation is evaluated on observations containing astrophysical signals and multiple interferers. The target source is PSR~J1713+0747 (spin period 4.5~ms, pulse width 1~ms, DM~=~15.97~$\mathrm{cm^{-3}\, pc}$) \cite{1993ApJ...410L..91F}. RFI includes Iridium at 1620–1626~MHz (channels~1402–1433), FAA radar at 1255 and 1305~MHz (channels 3302 and 3046), unidentified emitters near 1500~MHz (channel~2048), and GPS‑L3 at 1381~MHz (channel~2657). 

Eight observing epochs produced forty 1~GB files. Voltages were sampled at a 1600~MHz Nyquist rate and channelized via STFT into real/imaginary components across 4096 channels, 65,024 time samples, and two polarizations. The post‑channelization sampling interval is 5.12~$\mu\mathrm{s},$ giving ~0.332~s per file.

Each file is stored in MATLAB format as a 4096~$\times$~65,024 array (real and imaginary parts for both polarizations). This work analyzes one file.

 \subsection{Observational Results}
Both SK and the proposed RFI‑segmentation methods operate on spectrograms (power spectral densities) obtained by squaring the real and imaginary channelized voltages and summing them. We chose to work with a single polarization, leaving the other for future analysis. The resulting spectrogram is an array of size $K\times M=4096\times 65,024.$  

To apply SK to the spectrogram, each frequency channel $k,\  k=1,\ldots ,K,$ was divided into non‑overlapping intervals of length $512.$ Partitioning $65,024$ time samples into intervals of length $512$ yields $127$ intervals. The SK estimator (\ref{eq:generalizedSK}) is evaluated for each of the $127$ intervals, and its value is compared to the threshold $\pm \, 6/ \sqrt{M} \cdot \gamma,$ where $\gamma$ is a scale parameter varied between 3 and 7. If the value of the SK estimator exceeds the threshold, the corresponding $512$ time samples in that interval are set to zero; otherwise, they are retained. 

To apply the proposed RFI‑detection methods, we follow the RFI‑mitigation pipeline outlined in Section \ref{sec:methodology}. The length of STFT was set to $512$ in this experiment. 

\begin{figure}[htb]
    \centering
    \includegraphics[width=8.5cm]{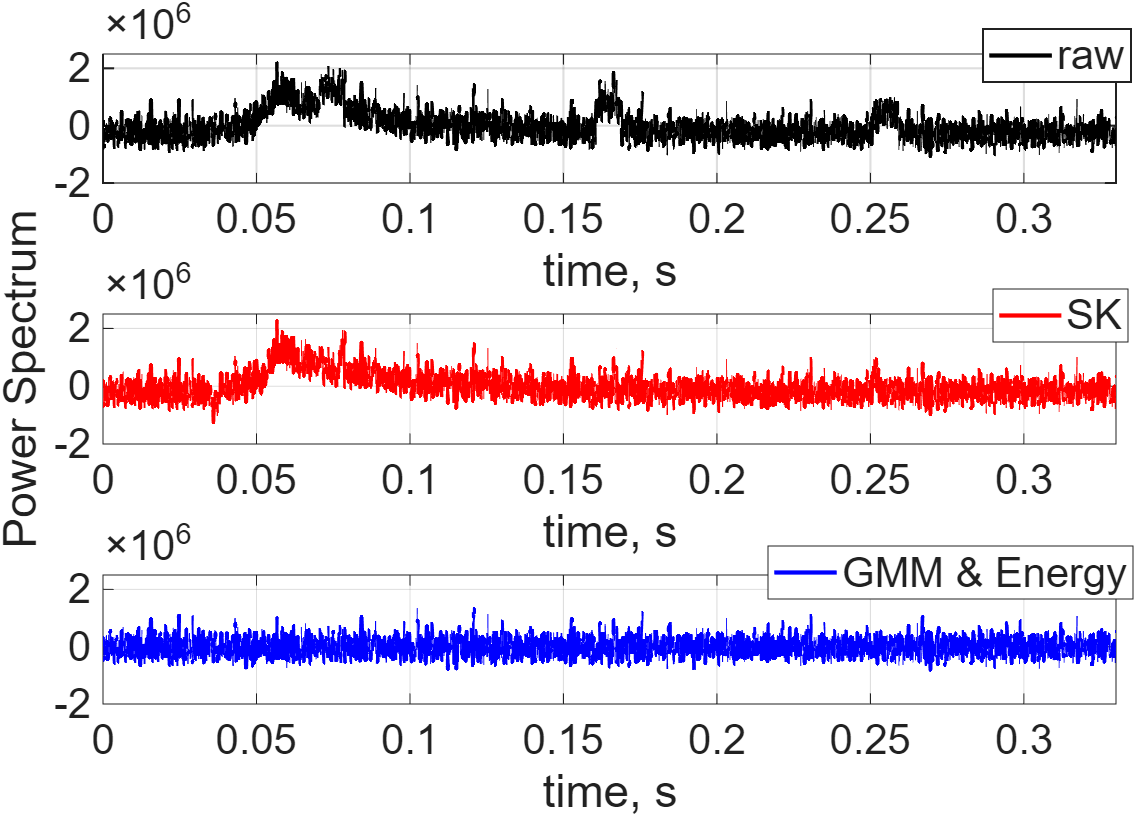}
    \caption{The power spectrum is presented as a time series containing 57 pulsar pulses embedded in noise and residual RFI.}
    \label{fig:Periodograms}
\end{figure}

To visually assess the effectiveness of the proposed pipeline, we display a periodogram cleaned of RFI, dedispersed at a DM of $15.97~\mathrm{cm^{-3}\, pc},$ and integrated in frequency for three cases: raw data (no RFI removal), data processed with SK, and data processed with the proposed methods. Figure \ref{fig:Periodograms} shows the results after applying the signal‑processing steps. Several pulses of J1713+0747 are clearly visible in each panel between 0.1 and 0.14 s.

To illustrate the range of RFI morphologies for which the proposed method is effective, in Figure \ref{fig:res} we show spectrograms from frequency channels 1436, 1818, 1830, and 1913 (left panels), together with their corresponding masks indicating regions labeled as RFI‑free (right panels). Regardless of whether the interference is narrowband, time‑varying, or densely structured, the segmentation algorithms successfully separate RFI from the background noise floor.
\begin{figure}[htb]
    \centering
    \includegraphics[width=8.5cm]{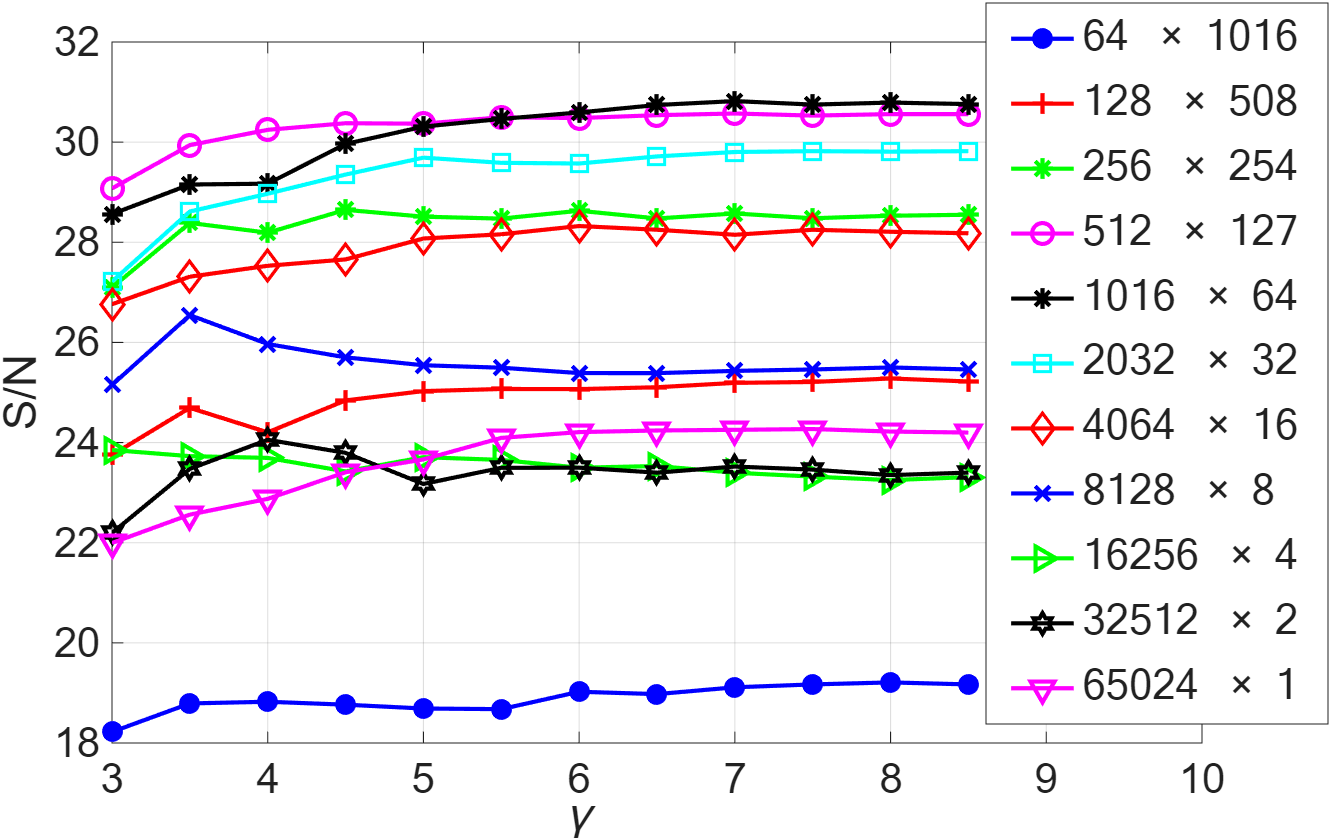}
    \caption{The effect of varying the STFT window length on the performance of GMM \& Energy thresholding method. }
    \label{fig:SNR_vs_Th}
\end{figure} 
\vspace{-0.1in}             

 \subsection{Qualitative Results}
To analyze the effectiveness of the newly proposed methods in the detection and mitigation of RFI, we compare their performance on the pulsar S/N against the performance of the SK method. Since the SK estimator and the MAD‑based method depend on the choice of the threshold parameter $\gamma$ we present the performance of the two approaches as a function of $\gamma.$  For the riptide \cite{Morello2020FFA} single‑pulse detection of PSR J1713+0747, the raw (unprocessed) data yield an S/N of 19.59. The SK method attains a maximum S/N of 22.85 at $\gamma = 5.0.$ Applying GMM‑based segmentation increases the S/N to 30.54. At last, the combined GMM and energy‑based thresholding method produces an S/N that remains effectively constant across a wide range of $\gamma,$ indicating robustness to the choice of thresholding parameter. 
\begin{table}
        \centering 
        \begin{tabular}{|c|c|c|c|c|c|} 
        \hline
        $\gamma$ & 3 & 4 & 5 & 6 & 7 \\
        \hline           
        \hline 
        SK & 20.77 & 22.19 & 22.85 & 21.93 & 20.95 \\
        \hline 
        GMM \& En. & 29.08 & 30.25 & 30.38 & 30.49 & 30.58 \\
        \hline          
        \end{tabular}
        \caption{This table presents the performance comparison of the proposed RFI mitigation methods, GMM \& Energy thresholding, with the performance of SK as the parameter $\gamma$ is varied from 3 to 7. }
        \label{table:pulseFreeData}
\end{table}                 
Our final experiment examines how varying the STFT window length affects the performance of the proposed method. As is well known, the window length governs the trade‑off between time and frequency resolution: longer windows provide higher frequency resolution at the cost of temporal resolution, whereas shorter windows improve temporal resolution but reduce frequency discrimination. Figure~\ref{fig:SNR_vs_Th} shows the S/N as a function of $\gamma,$ parameterized by the chosen window length $L.$ Each channel of the original spectrogram contained 65,024 time samples and was processed using STFT windows of length 64, 128, 256, 512, 1016, 2032, 4064, 8128, 32,512, and 65,024.

\section{Summary} 
\label{sec:summary}
%
We propose an RFI‑detection method for radio astronomy data based on applying the STFT to each frequency channel of the original spectrogram, where each discrete frequency corresponds to a band $\Delta W$ of width 195,312.5 Hz. Because RFI exhibits diverse temporal and spectral structure, subdividing $\Delta W$ into narrower subbands via the STFT improves localization at the cost of reduced time resolution. The STFT magnitude is treated as an image, and RFI is identified using GMM and an energy‑based MAD threshold, assuming RFI appears as high‑energy structured features. Detected components are nulled in the STFT domain, inverse‑transformed, and reinserted into the spectrogram. 

Performance is assessed using the S/N of a single folded pulse of PSR J1713+0747, compared against raw data and SK. STFT window lengths of 512 and 1024 applied to high‑resolution GBT spectrograms yield substantially higher S/N than SK, demonstrating that computer‑vision techniques effectively separate structured RFI from astronomical signals.

\clearpage     
\color{black} 
\bibliographystyle{IEEEbib}
\bibliography{frb_refs,refs}

\end{document}